\documentclass[preprint,tighten,aps]{revtex4}

\begin{document}

\baselineskip = 0.9\baselineskip
\oddsidemargin=10mm
\parindent=1cm

\title{Renormalized coupling constants for 3D scalar $\lambda\phi^4$
field theory and pseudo-$\epsilon$ expansion}

\author{M. A. Nikitina$^{1,2}$}
\author{A. I. Sokolov$^{1}$}
\email{ais2002@mail.ru}

\address
{$^{1}$Saint Petersburg State University, Saint Petersburg, Staryi Petergof,
Russia, \\ $^{2}$National Research University ITMO, Saint Petersburg, Russia}

\date{\today}

\begin{abstract}
{Renormalized coupling constants $g_{2k}$ that enter the critical equation
of state and determine nonlinear susceptibilities of the system possess
universal values $g_{2k}^*$ at the Curie point. They are calculated, along
with the ratios $R_{2k} = g_{2k}/g_4^{k-1}$, for the three-dimensional
scalar $\lambda\phi^4$ field theory within the pseudo-$\epsilon$ expansion
approach. Pseudo-$\epsilon$ expansions for $g_6^*$, $g_8^*$, $R_6^*$, and
$R_8^*$ are derived in the five-loop approximation, numerical estimates
are presented for $R_6^*$ and $R_8^*$. The higher-order coefficients of
the pseudo-$\epsilon$ expansions for the sextic coupling are so small that
simple Pad\'e approximants turn out to be sufficient to yield very good
numerical results. Their use gives $R_6^* = 1.650$ while the most recent
lattice estimate is $R_6^* = 1.649(2)$. For the octic coupling
pseudo-$\epsilon$ expansions are less favorable from the numerical point
of view. Nevertheless, Pad\'e--Borel resummation leads in this case to
$R_8^* = 0.890$, the number differing only slightly from the values $R_8^*
= 0.871$, $R_8^* = 0.857$ extracted from the lattice and field-theoretical
calculations.}

\vspace{1.0cm}


\end{abstract}

\maketitle

The critical behavior of the systems undergoing continuous phase transitions
and described by the three-dimensional (3D) Euclidean scalar $\lambda\phi^4$
field theory is characterized by a set of universal parameters including,
apart from critical exponents, renormalized effective coupling constants
$g_{2k}$ and their ratios $R_{2k}=g_{2k}/g_4^{k-1}$. These ratios enter the
small magnetization expansion of free energy and determine, along with
renormalized quartic coupling constant $g_4$, the nonlinear susceptibilities
of various orders:
\begin{equation}
F(z,m) - F(0,m) = {\frac{m^3 }{g_4}} \Biggl({\frac{z^2 }{2}} +
z^4 + R_6 z^6 + R_8 z^8 + ... \Biggr),
\end{equation}
\begin{eqnarray}
\chi_4 &=& {\frac{\partial^3M}{\partial H^3}} \Bigg\arrowvert_{H = 0} = -
24 {\frac{\chi^2}{m^3}} g_4, \\
\chi_6 &=& {\frac{\partial^5M}{\partial H^5}} \Bigg\arrowvert_{H = 0} =
- 6! {\frac{\chi^3 g_4^2}{m^6}}(R_6 - 8), \\
\chi_8 &=& {\frac{\partial^7M}{\partial H^7}} \Bigg\arrowvert_{H = 0} =
- 8! {\frac{\chi^4 g_4^3}{m^9}}(R_8 - 24 R_6 + 96),
\end{eqnarray}
where $z = M \sqrt{g_4/m^{1 + \eta}}$ is a dimensionless magnetization,
renormalized mass $m \sim (T - T_c)^\nu$ being the inverse correlation
length, $\chi$ is a linear susceptibility while $\chi_4$, $\chi_6$, and
$\chi_8$ are nonlinear susceptibilities of fourth, sixth, and eighth
orders.

For the 3D Ising model or, equivalently, for 3D scalar $\lambda\phi^4$ field
theory the nonlinear susceptibilities and the scaling equation of state are
intensively studied theoretically. During last four decades renormalized
coupling constants $g_{2k}$ and the ratios $R_{2k}$ were evaluated by a
number of analytical and numerical methods \cite{B77,LGZ77,BNM78,LGZ80,BB,
B1,B2,TW,B3,Reisz95,AS95,CPRV96,S96,ZLF,SOU,GZ97,Morr,BC97,GZJ98,BC98,PV98,
PV,S98,SOUK99,PV2000,ZJ01,CHPV2001,CPRV2002,ZJ,PV02,BP11,NS14}. Estimating
the universal critical values $g_4^*$, $g_6^*$ and $R_6^*$ by means
of the field-theoretical renormalization group (RG) approach in physical
dimensions has shown that RG technique enables one to get accurate
numerical estimates for these quantities. For example, four- and five-loop
RG expansions resummed with a help of Borel-transformation-based procedures
lead to the values for $g_6^{*}$ differing from each other by less than
$0.5\%$ \cite{SOU, GZ97} while the three-loop RG approximation turns out
to be sufficient to provide an apparent accuracy no worse than $1.6\%$
\cite{SOU,S98}. In principle, this is not surprising since the
field-theoretical RG approach proved to be highly efficient when used to
estimate critical exponents, critical amplitude ratios, marginal
dimensionality of the order parameter, etc. for numerous phase transition
models \cite{BNM78,LGZ80,AS95,GZJ98,ZJ01,ZJ,PV02,PS2000,CPV2000}.

To obtain proper numerical estimates from diverging RG expansions the
resummation procedures have to be applied. Most of those being used are
based upon the Borel transformation which kills the factorial growth of
higher-order coefficients and paves the way to converging iteration
schemes. These schemes have enabled to obtain a great number of accurate
numerical results for basic models of critical phenomena. There exists,
however, alternative technique turning divergent perturbative series into
more convenient ones, i. e. into expansions that have smaller lower-order
coefficients and much slower growing higher-order ones than those of
original series. The method of pseudo-$\epsilon$ expansion invented
by B. Nickel (see Ref. 19 in the paper of Le Guillou and Zinn-Justin
\cite{LGZ80}) is meant here. This approach has been shown to be very
efficient numerically when used to evaluate critical exponents and other
universal quantities of various 3D and 2D systems \cite{LGZ80,NS14,
FHY2000,HID04,CP05,COPS04,S2005,S2013,NS13,NS14e,NS15,NS16}.

In this paper, we study renormalized effective coupling constants and
universal ratios $R_{2k}$ of the 3D scalar $\lambda\phi^4$ field theory
with the help of pseudo-$\epsilon$ expansion technique. The
pseudo-$\epsilon$ expansions ($\tau$-series) for the universal values of
renormalized coupling constants $g_6$ and $g_8$ will be calculated on the
base of five-loop RG expansions obtained by Guida and Zinn-Justin
\cite{GZ97}. Along with the sextic and octic coupling constants, universal
critical values of ratios $R_6 = g_6/g_4^2$ and $R_8 = g_8/g_4^3$ will be
found as series in $\tau$ up to $\tau^5$ terms. The pseudo-$\epsilon$
expansions obtained will be processed by means of Pad\'e and Pad\'e--Borel
resummation techniques as well as by direct summation when it looks
reasonable. The numerical estimates for the universal ratios will be
compared with the values extracted from the higher-order
$\epsilon$-expansions, from the perturbative RG expansions in three
dimensions, and from the lattice calculations and some conclusions
concerning the numerical power of the pseudo-$\epsilon$ expansion approach
will be formulated.

So, the Hamiltonian of the model under consideration reads:
\begin{equation}
\label{eq:3} H = \int d^{3}x \Biggl[{1 \over 2}( m_0^2 \varphi_{\alpha}^2
 + (\nabla \varphi_{\alpha})^2)
+ {\lambda \over 24} (\varphi_{\alpha}^2)^2 \Biggr] ,
\end{equation}
where $\varphi_{\alpha}$ is a real scalar field, bare mass squared $m_0^2$
being proportional to $T - T_c^{(0)}$, $T_c^{(0)}$ -- mean field
transition temperature. The $\beta$-function for the model (3) has been
calculated within the massive theory \cite{Guelph,BNM78} with the propagator,
quartic vertex and $\varphi^2$ insertion normalized in a conventional way:
\begin{eqnarray}
\label{eq:4}
G_R^{-1} (0, m, g_4) = m^2 , \qquad \quad {{\partial G_R^{-1}
(p, m, g_4)} \over {\partial p^2}}
\bigg\arrowvert_{p^2 = 0} = 1 , \\
\nonumber
\Gamma_R (0, 0, 0, m, g) = m^2 g_4, \qquad \quad
\Gamma_R^{1,2} (0, 0, m, g_4) = 1.
\end{eqnarray}
Later, the five-loop RG series for renormalized coupling constants $g_6$
and $g_8$ of this model were obtained \cite{GZ97} and the six-loop
pseudo-$\epsilon$ expansion for the Wilson fixed point location was
reported \cite{NS14}:
\begin{eqnarray}
g_6 = {\frac 9\pi }g_4^3\Bigl(1-{\frac{3}{\pi}} g_4 +
1.38996295 g_4^2 - 2.50173246 g_4^3 + 5.275903\ g_4^4 \Bigr),
\end{eqnarray}
\begin{eqnarray}
g_8 = -{\frac{81}{{2\pi}}} g_4^4 \Bigl(1 - {\frac{65}{6\pi}}
g_4 + 7.77500131  g_4^2 - 18.5837685 g_4^3 + 48.16781 g_4^4\Bigr),
\end{eqnarray}
\begin{eqnarray}
g_4^* &=&\frac{2\pi}{9}\biggl(\tau + 0.4224965707\tau^{2} + 0.005937107
\tau^{3} + 0.011983594 \tau^{4}
\nonumber\\
&-& 0.04123101\tau^{5} + 0.0401346\tau^{6}\biggr).
\end{eqnarray}
Combining these expansions one can easily arrive to the $\tau$-series for
the values of the coupling constants $g_6$ and $g_8$ at the critical point:
\begin{eqnarray}
g_6^* = {8 \pi^2 \over 81}\tau^3 \bigl(1 + 0.600823045
\tau + 0.104114939 \tau^2 - 0.023565414 \tau^3 - 0.01838783 \tau^4 \bigr)
\end{eqnarray}
\begin{eqnarray}
g_8^* = -{8 \pi^3 \over 81}\tau^4 \bigl(1 - 0.717421125 \tau
- 0.201396988 \tau^2 - 0.70623903 \tau^3 + 0.8824349 \tau^4 \bigr)
\end{eqnarray}
Corresponding pseudo-$\epsilon$ expansions for the universal ratios
are as follows:
\begin{eqnarray}
R_6^* = 2\tau \bigl(1- 0.244170096 \tau + 0.120059430
\tau^2 - 0.1075143 \tau^3 + 0.1289821 \tau^4\bigr).
\end{eqnarray}
\begin{eqnarray}
R_8^* = -9\tau \bigl(1 - 1.98491084 \tau + 1.76113570
\tau^2 - 1.9665851 \tau^3 + 2.741546 \tau^4 \bigr).
\end{eqnarray}
These $\tau$-series will be used for evaluation of renormalized effective
couplings near the Curie point.

First, let us find the numerical value of the ratio $R_6$ at criticality.
Since the pseudo-$\epsilon$ expansion (12) has small higher-order
coefficients a direct summation of this series looks quite reasonable.
Within third, fourth and fifth orders in $\tau$ it gives 1.752, 1.537 and
1.795 respectively. These numbers certainly group around the estimates
1.644 and 1.649 extracted from advanced field-theoretical and lattice
calculations \cite{GZ97,BP11}. It is interesting that the value 1.537
obtained by truncation of the series (12) by the smallest term (optimal
truncation \cite{NS13}) differs from the estimates just mentioned by 6\%
only. Moreover, direct summation of $\tau$-series for $g_6^*$ (10) having
very small higher-order coefficients gives the value $g_6^* = 1.621$ which
under $g_4^* = 0.9886$ \cite{BNM78} results in the estimate $R_6^* =
1.659$ looking rather optimistic. This fact confirms the conclusion that
the pseudo-$\epsilon$ expansion itself may be considered as some specific
resummation method \cite{NS14,NS13,NS14e,NS15,NS16}.

Much more accurate numerical value of $R_6^*$ can be obtained from the
pseudo-$\epsilon$ expansion (12) using Pad\'e approximants [L/M]. Pad\'e
triangle for $R_6^*/\tau$, i. e. with the insignificant factor $\tau$
neglected is presented in Table I. Along with the numerical values given
by various Pad\'e approximants the rate of convergence of Pad\'e estimates
to the asymptotic value is shown in this Table (the lowest line, RoC).
Since the diagonal and near-diagonal Pad\'e approximants are known to possess
the best approximating properties the Pad\'e estimate of $k$-th order is
accepted to be given by the diagonal approximant or by the average over
two near-diagonal ones when corresponding diagonal approximant does not
exist. As seen from Table I, the convergence of Pad\'e estimates is well
pronounced and the asymptotic value of $R_6^*$ equals 1.6502. This number
is close to the higher-order $\epsilon$ expansion estimate
$R_6^* = 1.690 \pm 0.04$ \cite{GZ97}, to the five-loop 3D RG estimate
$R_6^* = 1.644 \pm 0.006$ \cite{GZ97} and, in particular, to the
value $R_6^* = 1.649 \pm 0.002$ given by the advanced lattice
calculations \cite{BP11}.

It is worthy to note that an account for the five-loop terms in
$\tau$-series for $R_6^*$ and $g_6^*$ shifts the numerical value of the
universal ratio only slightly. Indeed, the Pad\'e resummation of the
four-loop $\tau$-series for $R_6^*$ and $g_6^*$ result in $R_6^* = 1.642$
and $R_6^* = 1.654$, respectively \cite{NS14}. This may be considered as a
manifestation of the fact that the $\tau$-series for the sextic effective
coupling have a structure rather favorable from the numerical point of
view. It is especially true for the pseudo-$\epsilon$ expansion (12)
which, having small higher-order coefficients, is alternating what makes
this series very convenient for getting numerical estimates.

For the renormalized octic coupling we have pseudo-$\epsilon$ expansions
with much less favorable structure. The series for $R_8^*$ (13) being
alternating possesses rather big coefficients making a direct summation
certainly useless in this case. To estimate the ratio $R_8^*$ Pad\'e
resummation procedure should be applied. The coefficients of the
$\tau$-expansion for $g_8^*$ (11) are considerably smaller than those of
the series (13) but have irregular signs. This series may be also
processed, in principle, within the technique mentioned. The numerical
results thus obtained, however, turn out to be so strongly scattered that
they could not be referred to as satisfactory or even meaningfull. That is
why further we will concentrate on the Pad\'e resummation of the
$\tau$-series for $R_8^*$.

We construct Pad\'e approximants for the ratio $R_8^*/\tau$ neglecting, as
in the case of $R_6^*$, the insignificant factor $\tau$. Corresponding
Pad\'e triangle is presented in Table II. As is seen, in the case of octic
coupling numerical estimates turn out to be much worse than those obtained
for $R_6^*$. Indeed, the numbers given by Pad\'e resummed
pseudo-$\epsilon$ expansion (13) are strongly scattered. Moreover, even
the estimates extracted from the highest-order available -- five-loop --
$\tau$-series differ from each other considerably. On the other hand, the
optimal value of $R_8^*$, i. e. that given by the diagonal approximant
[2/2] is close to the 3D RG estimate $R_8^* = 0.857 \pm 0.086$ \cite{GZ97}
and to the result of recent lattice calculations $R_8^* = 0.871 \pm 0.014$
\cite{BP11}. It implies that use of more powerful resummation techniques
may lead to acceptable results.

One of such techniques effectively suppressing divergence of the series
being resummed is the Pad\'e--Borel machinery. The results of resummation
of pseudo-$\epsilon$ expansion (13) with the help of this method are
presented in Table III. As is seen, use of Pad\'e--Borel technique makes
numerical estimates much less scattered and markedly accelerates the
iteration procedure. Moreover, the asymptotic value 0.890 the iterations
result in only slightly differs from the high-precision estimates
mentioned above. This shows that the pseudo-$\epsilon$ expansion technique
remains workable when employed to estimate the universal value of the
higher-order coupling constants.

At the same time, the question arises: what is the origin of the poor
approximating properties of the $\tau$-series for $g_8^*$ and $R_8^*$? Of
course, pronounced divergence of these pseudo-$\epsilon$ expansions may be
thought of as a main source of such a misfortune. There exists, however,
an extra moment making the situation rather unfavorable. The point is that
the series (11), (13) have some specific feature. Namely, their first
terms are negative and big in modulo while the estimates these series
imply turn out to be an order of magnitude smaller and have an opposite
sign. It means that numerical values resulting from (11), (13) are nothing
but small differences of big numbers what drastically lowers an
approximating power of these $\tau$-series.

So, we have calculated pseudo-$\epsilon$ expansions for the universal
values of renormalized coupling constants $g_6$, $g_8$ and of the ratios
$R_6$, $R_8$ for 3D Euclidean scalar $\lambda\phi^4$ field theory.
Numerical estimates for $R_6^*$ and $R_8^*$ have been found using
Pad\'e and Pad\'e--Borel resummation techniques. The pseudo-$\epsilon$
expansion machinery has been shown to lead to high-precision value of
$R_6^*$ which is in very good agreement with the numbers obtained by
means of other methods including lattice calculations. For the octic
coupling this technique has been found to be less efficient. However
the numbers extracted from $\tau$-series for $R_8^*$ by means of the
Pad\'e--Borel resummation turn out to be rather close to high-precision
lattice and field-theoretical estimates. These results confirm a
conclusion that the pseudo-$\epsilon$ expansion approach may be referred
to as a specific resummation method converting divergent RG series into
the expansions very convenient from the numerical point of view.

We gratefully acknowledge the support of Saint Petersburg State University
via Grant 11.38.636.2013, of the Russian Foundation for Basic Research
under Project No. 15-02-04687, and of the Dynasty foundation.

\newpage

\begin{table}[t]
\caption{Pad\'e table for pseudo-$\epsilon$ expansion of the ratio
$R_6^*$. Pad\'e approximants [L/M] are derived for $R_6^*/\tau$, i. e.
with factor $\tau$ omitted. The lowest line (RoC) shows the rate of
convergence of Pad\'e estimates to the asymptotic value. Here the Pad\'e
estimate of $k$-th order is that given by diagonal approximant or by the
average over two near-diagonal ones when corresponding diagonal
approximant does not exist. The values of $R_6^*$ resulting from the 3D RG
analysis \cite{GZ97} and recent lattice calculations \cite{BP11} are equal
to $1.644 \pm 0.006$ and $1.649 \pm 0.002$ respectively.}
\label{tab1}
\renewcommand{\tabcolsep}{0.4cm}
\begin{tabular}{{c}|*{5}{c}}
$M \setminus L$ & 0 & 1 & 2 & 3 & 4  \\
\hline
0 & 2      & 1.5117 & 1.7518 & 1.5368 & 1.7947 \\
1 & 1.6075 & 1.6726 & 1.6383 & 1.6540  \\
2 & 1.6896 & 1.6465 & 1.6502 &  \\
3 & 1.6036 & 1.6504  \\
4 & 1.7135 \\
\hline
RoC & 2    & 1.5596 & 1.6726 & 1.6424 & 1.6502 \\
\end{tabular}
\end{table}

\begin{table}[t]
\caption{Pad\'e triangle for pseudo-$\epsilon$ expansion of the ratio
$R_8^*$. Pad\'e approximants [L/M] are constructed for $R_8^*/\tau$, i. e.
with factor $\tau$ omitted. The lowest line (RoC) demonstrates the rate of
convergence of Pad\'e estimates where Pad\'e estimate of $k$-th order is
that given by diagonal approximant or by the average over two
near-diagonal ones if diagonal approximant does not exist.}
\label{tab2}
\renewcommand{\tabcolsep}{0.4cm}
\begin{tabular}{{c}|*{5}{c}}
$M \setminus L$ & 0 & 1 & 2 & 3 & 4 \\
\hline
0   & $-9$     & 8.864  & $-6.986$ & 10.713   & $-13.961$ \\
1   & $-3.015$ & 0.466  & 1.376    & 0.407 \\
2   & $-1.743$ & 1.910  & 0.879    \\
3   & $-1.131$ & 0.095  \\
4   & $-0.831$ \\
\hline
RoC & $-9$     & 2.925  & 0.466    &  1.643   & 0.879 \\
\end{tabular}
\end{table}

\begin{table}[t]
\caption{The values of universal ratio $R_8$ obtained by means of Pad\'e--Borel
resummation of the series (13). Pad\'e approximants [L/M] are employed for
analytical continuation of the expansion of Borel transform of $R_8^*/\tau$
in powers of $\tau$. The lowest line (RoC) demonstrates the rate of convergence
of Pad\'e--Borel estimates to the asymptotic value. The estimate of $k$-th
order is that found using corresponding diagonal approximant or the average
over two values given by approximants [M/M$-$1] è [M$-$1/M] when the diagonal
approximant does not exist. The values of $R_8^*$ resulting from 3D RG analysis
\cite{GZ97} and extracted from advanced lattice calculations \cite{BP11} are
equal to $0.857 \pm 0.086$ and $0.871 \pm 0.014$ respectively.}
\label{tab3}
\renewcommand{\tabcolsep}{0.4cm}
\begin{tabular}{{c}|*{5}{c}}
$M \setminus L$ & 1 & 2 & 3 & 4 & 5 \\
\hline
0 & $-9$      & 8.8642  & $-6.9860$ & 10.7132  & $-13.9607$ \\
1 & $-3.6473$ & 1.1000  & 0.8854  & 0.8906 \\
2 & $-2.4658$ & 0.9041  & 0.8905  \\
3 & $-2.0048$ & 0.8916 \\
4 & $-1.7788$ \\
\hline
RoC &  $-9$   & 2.608   &  1.1000  & 0.8948  & 0.8905 \\
\end{tabular}
\end{table}

\end{document}